# Calculations of the Efficiency of Registration of Thermal Neutrons by Complex Converters Constructed on the Basis of Gadolinium Foils

D. A. Abdushukurov[*)], D.V.Bondarenko, Kh.Kh.Muminov[**)], D.Yu.Chistyakov

Physical-Technical Institute of Academy of Sciences of the Republic of Tajikistan, Aini Ave 299/1, Dushanbe 734063, Tajikistan

e-mails: [*)] abdush@tajik.net, [**)] muminov@tascampus.eastera.net

Abstract: We consider the results of modeling of the efficiency of registration of thermal neutrons by the converters, which are made from natural gadolinium and its 157 isotope foils. Efficiency for a case of falling of neutrons under various angles to a plane of converters is calculated. It is shown, that at small angles of falling of neutrons to a plane of converters it is possible to receive the efficiency of registration close to a theoretical limit. Efficiency of the complex converter made of kapton supporting film with gadolinium converters layered on both its sides is considered. All calculations are carried out for four fixed neutron energies, which correspond to the wavelengths of 1, 1.8, 3 and 4 $A^0$.

## 1. Introduction

Among the solid-state converters of thermal neutrons the highest efficiency of registration can be received at use of converters on the basis of gadolinium and especially its 157 isotope. In the paper /1,2/ modeling of the efficiency of registration of thermal neutrons by converters made from gadolinium foils has been carried out. The following conditions were considered at this modeling: the neutron beam (with 2 fixed energies) perpendicularly fell on the converting foils of a various thicknesses, the secondary electron escapes were calculated for the both cases of emission, i.e. to the forward and to the back hemispheres.

Recently many authors were engaged in design and manufacturing of the detectors on the basis of gadolinium converters /3,8/. Modeling of the efficiency of registration of thermal neutrons /9,12/ is also conducted. This testifies to the indefatigable interest to gadolinium converters.

It is worth to mention, that modeling of the processes of neutron capture by gadolinium nuclei is also used in radiation therapy, in order to decrease doses of radiation received by oncological patients. Recently, since 2000, the new branch of oncology and treatment of cancer tumours, namely, a gadolinium neutron-capture therapy (GdNCT) /13/ has started to develop. The method is based on introduction into an organism of gadolinium nuclei as a part of medical products and their selective absorption by the cells of malignant tumour. Thus gadolinium nuclei, possessing an extraordinary high section of interaction with thermal neutrons, allow to localize influence of radiation by the area of cancer cells. Therefore the basic radiating influence is rendered by both low-energy electrons of internal conversion and Auger electrons.

In this paper we discuss the results of our modeling of efficiency of registration of thermal neutrons by the converters made of natural gadolinium and its 157 isotope foils. In our early works /11,12/ we presented the results of calculations of efficiency of gadolinium converters in the case when a neutron beam falls perpendicularly to a converter plane. Model representation, procedure of calculations and the received results are described in the above cited papers. Search of new ways of use of converters is needed in order to increase the efficiency of registration of neutrons and construction of detectors of a big area. One of the possible ways here is the use of converters located under the small angles to a plane of falling of neutron beams. During the construction of detectors of big areas the developers face a following problem (especially it concerns to the case of use of 157 isotope of gadolinium). In order to increase the efficiency of the detector the thickness of the converter should be about 0.5-3 microns, however the foils with such a thickness do not possess a sufficient mechanical durability and it requires using of a supporting film made from kapton or similar materials. Results of modeling of these kind converters are also addressed in this paper.

In this paper we present the results of modeling of the efficiency for the case of falling of neutrons under various angles to the plane of converters. We show that at small angles of falling of neutrons to the plane of converters it is possible to receive the efficiency of registration close to its theoretical limit.

In the second part of the paper we address the efficiency of the complex converter consisting of a supporting kapton film with a gadolinium converters layered on its surface from the both sides. All calculations are performed for four fixed neutron energies, which correspond to the wavelengths of 1, 1.8, 3 and 4 A$^0$.

2. Physical bases

In the process of radiating capture of thermal neutrons by gadolinium nuclei besides radiation of $\gamma$ - quanta also electrons of internal conversion and Auger electrons are emitted. Basically these electrons are registered by the position-sensitive detectors and especially by the gas-filled ones, which have a low efficiency of registration of $\gamma$ - quanta. Therefore, in the course of calculations we have considered only those electrons, which appeared as a result of conversion of neutrons on the gadolinium nuclei.

Natural gadolinium consists of the mix of isotopes, which are able to take part in the reaction of radiating capture of neutrons (n, $\gamma$) /14/. Note that for our calculations only natural gadolinium and its 155 and 157 isotopes, which have abnormal high sections of interaction with neutrons, are of special interest. Other isotopes bring an insignificant contribution into the interaction with neutrons.

In the reaction of neutron capture by $^{157}$Gd totally 7937.33 keV of energy is released. In total 390 lines with the energy range from 79.5 up to 7857.670 keV with the intensities of lines from $2 \cdot 10^{-3}$ up to 139 gamma-quanta by 100 captured neutrons /15/ are emitted.

As there are low-energy gamma- quanta in the spectrum, therefore in the process of their emission electrons from a nuclear shell (electrons of internal conversion) are irradiated with a high probability. The nucleus removes its excitation by irradiation of a gamma-quantum, but it also can emit a close located electron. As usual K-electron (an electron from K-shell) is emitted, but there also could be emitted electrons from the higher shells (L, M, N and so on). Vacancy of an electron (an electron hole), formed as a result of this process, is filled by another electron from a higher level. This process is accompanied by radiation of X-ray quantum, or by radiation of Auger electron.

The effect of internal conversion is accompanied by a considerable X-ray radiation that can have positive influence at the use of the solid-state gadolinium based detectors. In our calculations we considered only those electrons, energy of which is higher than 20 keV. In total we considered 444 the most intensive discrete energies of electrons with the probability of their emission higher than $10^{-5}$ on 100 falling neutrons /16/. Fig. 1 shows the histogram of dependence of intensity of electrons on their energies.

The most important characteristic of the converter is the probability of escape of secondary electrons, emitted in the process of radiating capture, from a converter material. As the most electrons possess low energies and have short runs in a converter material, it puts forward additional requirements to a thickness of the converter.

The coefficient of absorption $F_0$ characterizes the probability of absorption of electrons in a substance. If X is a thickness of the converter, Re is the run of electrons in the converter material, therefore the coefficient of absorption is defined by the following equation

$$F_0(X) = 1 - X \rho / Re, \qquad (1)$$

here $\rho$ is the density of the converter material. For gadolinium it is equal to $\rho$ = 7.9 g/cm$^3$.

The value of Re is determined by the energy of an electron and the value of the specific ionization losses,

$$Re = \int_0^E dE(-dE/dX) . \qquad (2)$$

Attenuation of the narrow collimated neutron beam by a thin layer of a substance is governed by the exponential law

$$F_x = F_0 \exp(-N_A \sigma X) \qquad (3)$$

where $F_x$ and $F_0$ are the densities of a neutron beam after and before of it passage through the layer of a substance having the thickness of X, accordingly, $N_A$ is the number of nuclei in 1 cm$^3$, $\sigma$ is full microscopic section interaction of neutrons with substance nuclei.

Calculations have been carried out for four fixed energies of neutrons, which correspond to the wavelengths of 1, 1.8, 3 and 4 $A^0$.

### 3. Calculations for plane-parallel foils

We started our modeling from consideration of a simple shape of target, i.e. first we considered converters constructed by plane-parallel foils. Calculations were conducted for thermal neutrons with fixed energies, corresponding to the neutron wavelengths of 1, 1.8, 3 and 4 $A^0$, the thickness of converters was varied from 1 to 40 microns, and also an isotope composition of the converter was varied, i.e. we considered converting foils made of natural gadolinium and its 157 isotope.

In the course of our calculations we have taken into consideration all electrons (appeared as a result of act of neutron capture), which are able to escape an infinite plain-parallel plate of the converter. Thus the ratio of the number of electrons escaped from a plate to the number of falling neutrons we referred to as the efficiency of the converter.

Conventionally we divide a thickness of a foil into more thin components. For each elementary layer we count the probability of absorption of neutrons with the fixed energy. Efficiency of the converter will be defined by the sum of probabilities of absorption of neutrons and probability of escape of electrons from the converter body. In order to calculate an electron escape we have chosen a simple model, namely, a geometrical one. The choice of the model is made taking into account the following assumptions: all electrons are emitted isotropic, the run length $Re_i$ for the each fixed electron energy is constant (we neglect the fluctuation of energy losses by the end of run).

Then the probability density to find an electron in the converter material forms a sphere with the radius $Re_i$ (for the each fixed energy). If we cross the center of the sphere by a plane, then two identical hemispheres, which correspond to electron emission to a forward and a backward hemispheres are formed, thus we consider the area of hemispheres as the probability of an electron escape. In this case the probability is 100%, and escape to one of hemispheres is 50 %. If we cross a sphere by planes with the step much less than $Re_i$, then there will be formed the segments with the areas, which are equal to the probability of electron escape. The step of iterations should be at least 100 times less than $Re_i$, and then it is possible to neglect the absorption of electrons, which are emitted under the big angles. If we cross a sphere by a plane at the distance $Re_i$, the area of the segment and, accordingly, the probability of electron escape will be equal to zero. The sum of probabilities of the electron escape for all energies, taking into account their weight contribution, determines the total probability of electron escape.

Released energies of electrons conventionally could be divided into 4 groups (see Fig. 1). The first one is formed by the least energetic electrons having small runs in a substance, and the fourth group is formed by the most energetic electrons.

We calculated the escape probability of isotropic emitted electrons for gadolinium. Fig. 2 shows contributions of electrons of different energetic groups. Calculations were made for conventionally fixed point of conversion of neutrons, which have co-ordinates (X=0, Y=0). One can easily see from the Fig.2 the contributions of each of 4 groups of electrons, which differ by their energies.

In our calculations we determined the probability of absorption of neutrons for each elementary layer as a result of non-elastic interaction $(n, \gamma)$ by use of the database created for fixed energies of neutrons. We calculated the probability of electron emission and the probability of electron escape from the converter body.

We have obtained the data on the efficiency of converters and their optimal thickness, for natural gadolinium and its 157 isotope (see tables 1 and 2 accordingly).

Table 1

| $\lambda$ | Efficiency E and | Optimal Thickness | (T [$\mu$ m]) |
|---|---|---|---|
| | forward | backward | total |
| 1 | 0.086 (13) | 0.107 (20) | 0.190 (17) |
| 1.8 | 0.145 (9) | 0.196 (18) | 0.326 (12) |
| 3 | 0.174 (6) | 0.234 (15) | 0.388 (9) |
| 4 | 0.190 (5) | 0.254 (13) | 0.425 (6) |

Table 2

| $\lambda$ | Efficiency E and | Optimal Thickness | (T [$\mu$ m]) |
|---|---|---|---|
| | forward | backward | total |
| 1 | 0.215 (4) | 0.272 (10) | 0.459 (6) |
| 1.8 | 0.256 (3) | 0.296 (8) | 0.532 (3) |
| 3 | 0.276 (2) | 0.317 (6) | 0.577 (3) |
| 4 | 0.288 (2) | 0.324 (4) | 0.605 (2) |

On the Fig. 3 and Fig. 4 the dependences of the efficiency of registration of neutrons with different energies on the thickness of the converter are shown.

We compared the accuracy of our calculations with the experimental data presented in the paper /10/ (G.Bruckner et al.). Data are received on the reactors of Atominstitut in Vienna (ATI) and ILL in Grenoble. Authors of the paper /10/ measured the efficiency of gadolinium converters in the geometry of escape of electrons in a back hemisphere and compared with the test $^3$He counter. Authors of this work used converters, which were made of natural gadolinium and its 157 isotope enriched up to 90.5 %.

Results of our modeling (see Fig. 5) are in agreement with the experimental results /10/ for natural gadolinium within limits of errors and give a little bit more low value for 157 isotope of gadolinium. Errors of calculations are caused by the accuracy in determining of gamma-quanta emission and errors in determining of section of interaction of neutrons. For the 157 isotope of gadolinium the experimental results /10/ are a little above the calculated data. Our data are in a good agreement with a theoretical limit, which is determined by the highest possible coefficient of internal conversion, and which defines the probability of formation of electrons of internal conversion and Auger electrons as 64 % for isotropic emitting electrons and as 32 % for escape of electrons to one of hemispheres.

## 4. Calculations of efficiency depending on angles of falling of neutrons on a foil

In the course of the further calculations we varied angles of falling of neutrons onto a foil. Having the fixed thickness, one can increase considerably the length of run of neutrons in a converter material at falling of neutrons under small angles on a foil plane, at the same time the path for escape of secondary electrons will remain short and constant. The direction of a neutron flux along the converter of a small thickness would be the most ideal case. This case could not be probably implemented in practice, but at the same time it allows one to estimate theoretical or maximal possible efficiency.

On the Fig. 6 the data on calculations of dependence of efficiency on the angles of falling of neutrons for $^{157}$Gd for two different thicknesses of converters (1 and 3 microns) are presented. The picture shows, that at a small thickness of the converter (1 micron and less) an insignificant difference at electron escapes in forward and backward hemispheres is observed. The maximum probability of an electron escape makes approximately 32 and 35 % for escape of electrons to the forward and the backward hemispheres, accordingly. The total efficiency reaches 65 % that could be considered as the maximal possible efficiency for gadolinium converters. The difference becomes more essential at increase of the thickness of converters.

We conducted the same calculations for converters made from $^{nat}$Gd as well (see Fig. 7). From this figure one can see, that at extremely small angles of falling of neutrons (2-3$^0$) the converters made from natural gadolinium are able to compete with the converters made from $^{157}$Gd.

We have also conducted the calculations for a case of falling of neutrons under the angle of $10^0$ for various thicknesses of converters for the neutron wavelength of $1.8A^0$ for two types of converters, i.e. made from 157 isotope and natural gadolinium, see Fig. 8. At the use of converters made from $^{157}$Gd with the thickness of 1 micron an efficiency of 64% could be reached, which could be considered as a theoretical limit.

### 5. Calculations of converters consisting of supporting films with thin converters layered on their surfaces

Essentially new constructional design decisions are necessary for manufacturing of thin converters (1-5 microns) as such thin foils do not possess a sufficient mechanical durability and there is not possible to construct the gas-filled detectors of the big sizes on their basis. One of the possible ways of the solution of this problem is the detector design which was offered in the papers /6,7/. In this design the converters are used as layers on the both surfaces of a supporting film. The supporting film is made from kapton or similar materials. This design uses two detectors which are located on the both sides of the converter. As the detectors there could be used various gas avalanche detectors, such as gas avalanche detectors of normal or low pressure. A good spatial resolution could be received by the use of the detectors of normal pressure. But in the case of use of the detectors of low pressure there is necessary to use the secondary emitters of electrons, which are made on the basis of CsI /8/. It is caused by the big value of length of free runs of secondary electrons, which can worsen the spatial resolution.

Let us consider in more detail the process of secondary electron registration in the detectors with such a complex converter. At an arrangement of the converters from the both sides of a supporting film, the part of neutrons will be converted in the first converter and a part in the second one. From the first converter electrons escaping into a back hemisphere will be registered, note these electrons will be registered by the forward chamber (by the 1 detector), see Fig. 9. From the second converter electrons escaping into a forward hemisphere will be registered, and these electrons will be registered accordingly by the back chamber (by the 2 detector). These processes will be the primary ones. At the same time a part of high energetic electrons can penetrate through the supporting film and could be also registered by the next chamber. These processes could increase a little the total efficiency of registration.

We calculated attenuation of a stream of electrons in gadolinium and kapton, see Fig. 10. The data on the spectrum of electrons irradiated in the process of radiating capture of neutrons by gadolinium nuclei were used at these calculations. One can see from the Fig. 10 that at the thickness of kapton film in 50 microns there is an appreciable reduction of probability of an escape of secondary electrons.

As a model we consider the case of an arrangement of gadolinium foil of the thickness of 1 micron, which was directly layered on the kapton films of a various thickness. Calculations are made for 4 different energies of neutrons corresponding to the lengths of waves of 1, 1.8, 3 and 4$A^0$. Features of variations of curves neither depend on isotope composition of converters nor on energy of neutrons, but they only depend on the thickness of an absorbing material. Especially low-energetic electrons are absorbed at a high rate. Results of these calculations are shown on the Fig. 11.

We calculated the possible efficiency of registration by the foils of a various thicknesses (1-3 microns) without use of supporting films. At the use of converters made from $^{157}$Gd the optimal thickness for an electron escape in a backward hemisphere is the thickness of 1-2 microns, see Fig. 12. Up to the thickness of 3 microns the growth of the efficiency is observed, however this converter is the lobby one and the further increase in its thickness leads to the shielding of the second converter. For the second converter (an electron escape in a forward hemisphere) the optimal thickness is 1-2 microns as the further increase in its thickness does not lead to increase of the efficiency.

 Further we have carried out calculations of the efficiency of the complex converters consisting of two converting foils located from the both sides of a supporting foil. Calculations were made for the following of geometry of converters, - thicknesses of the converters were chosen as 1+1 microns, 2+2 microns and 3+3 microns, converters are directly layered onto the surfaces of the kapton films with the thicknesses of 50 microns. The contribution of each converter to the efficiency of a separately taken detector and their total efficiency were calculated. Attenuation of the electron streams during their passage through the supporting films has been taken into consideration.

Results of calculations for the case of the converters made from $^{157}$Gd are presented on the Fig. 13. Fig. 14 shows the results of the same calculations, which were conducted for the converters made from $^{nat}$Gd. From these figures one can easily see, that in the case of use of 157 isotope of gadolinium the basic absorption of neutrons occurs on the lobby converter. The second converter is substantially shielded by the first one and at the thickness of converters more than 3 microns there the necessity for use of the second converter loses its meaning. Another picture is observed for the converters made from natural gadolinium, for this kind of converters the thickness should not be less than 3 microns.

In the Table 3 values of contributions of each of both converters made from $^{157}$Gd in the efficiency of the back detector, an electron escape in a forward hemisphere, for neutron wavelength of 1.8 $A^0$ are presented. Total efficiency of the back converter (the second detector) falls with the increase in its thickness, it happens due to its shielding by the first converter.

Table 3

| Geometry | The contribution of converters to efficiency (forward electron escape) of the second detector | | | Total efficiency |
|---|---|---|---|---|
| | first layer | Output from the 1 layer in view of attenuation | second layer | Total for two converters |
| 1 + 1 microns | 0.160 | 0.021 | 0.074 | 0.095 |
| 2 + 2 microns | 0.208 | 0.027 | 0.045 | 0.072 |
| 3 + 3 microns | 0.202 | 0.026 | 0.041 | 0.067 |

Table 4 shows the values of contributions of each of both converters made from $^{157}$Gd in the efficiency of the forward detector, an electron escape in a back hemisphere is considered. Neutron wavelength is 1.8 A$^0$.

Table 4

| Geometry | The contribution of converters to efficiency (back electron escape) of the first detector | | | Total efficiency |
|---|---|---|---|---|
| | first layer | second layer | Output from the 2 layer in view of attenuation | Total for two converters |
| 1 + 1 microns | 0.161 | 0.062 | 0.008 | 0.169 |
| 2 + 2 microns | 0.223 | 0.031 | 0.004 | 0.227 |
| 3 + 3 microns | 0.246 | 0.011 | 0.0015 | 0.247 |

The total efficiency of the first detector increases with the increase in the thickness of the converter.

In the Table 5 results of calculations conducted for each of detectors in the case of use of the converters made from $^{157}$Gd, for the neutron wavelengths of 1, 1.8, 3 and 4 A$^0$, are presented.

| Geometry, microns | Efficiency of registration by the second detector | | | | Efficiency of registration by the first detector | | | |
|---|---|---|---|---|---|---|---|---|
| | $1A^0$ | $1.8A^0$ | $3A^0$ | $4A^0$ | $1A^0$ | $1.8A^0$ | $3A^0$ | $4A^0$ |
| 1 + 1 | 0.056 | 0.095 | 0.126 | 0.084 | 0.065 | 0.169 | 0.210 | 0.235 |
| 2 + 2 | 0.076 | 0.072 | 0.037 | 0.034 | 0.106 | 0.227 | 0.26 | 0.277 |
| 3 + 3 | 0.076 | 0.067 | 0.034 | 0.029 | 0.130 | 0.248 | 0.273 | 0.285 |

The total efficiency of two detectors, providing registration of electrons in both directions, accordingly will make 0.26, 0.30 and 0.32, for double converters, which are made from $^{157}$Gd with the thicknesses 1, 2 and 3 microns. Calculations are presented for the neutron wavelength of 1.8 $A^0$.

| Geometry | Total efficiency of both detectors $^{157}$Gd | | | |
|---|---|---|---|---|
| | $1A^0$ | $1.8A^0$ | $3A^0$ | $4A^0$ |
| 1 + 1 microns | 0.12 | 0.26 | 0.34 | 0.34 |
| 2 + 2 microns | 0.18 | 0.30 | 0.30 | 0.31 |
| 3 + 3 microns | 0.21 | 0.32 | 0.30 | 0.31 |

The total efficiency of two detectors, which provide registration of electrons in both directions, accordingly will make 0.08, 0.14 and 0.16 for the double converters made from $^{nat}$Gd with the thicknesses 1, 2 and 3 microns. Calculations are carried out for the neutron wavelength of 1.8 $A^0$.

| Geometry | Total efficiency of both detectors $^{nat}$Gd | | | |
|---|---|---|---|---|
| | $1A^0$ | $1.8A^0$ | $3A^0$ | $4A^0$ |
| 1 + 1 microns | 0.026 | 0.084 | 0.116 | 0.140 |
| 2 + 2 microns | 0.046 | 0.135 | 0.175 | 0.202 |
| 3 + 3 microns | 0.060 | 0.160 | 0.200 | 0.218 |

Conclusion

In this paper we considered the results of modeling of the efficiency of registration of thermal neutrons by the converters made of foils of natural gadolinium and its 157 isotope. In the case of falling of a flux of neutrons perpendicularly to a converter plane it is possible to receive the efficiency of registration 0.17, 0.23 and 0.39, for the cases of an electron escape in a forward, a back hemispheres and a total efficiency, accordingly, for the converters made from natural gadolinium. Results of our calculations show, that the optimal thickness of the converter is 6 microns for the escape of electrons in a forward hemisphere and 15 microns for the escape in a back hemisphere. At the use of 157 isotope of gadolinium it is possible to receive the efficiency of registration 0.26, 0.30 and 0.53, accordingly, for the cases of an electron escapes in a forward, a back hemispheres and the total value of efficiency. The optimal thickness of the converter will make 3 microns for an electron escape in a forward hemisphere and 8 microns in the case of escape in a back hemisphere.

The increase in the efficiency of registration of neutrons can be reached at use of converters located under the small angles to a plane of falling of neutron fluxes. Results of modeling of these kinds of converters are presented also in this paper. Efficiency for a case of falling of neutrons under various angles to a plane of converters is calculated. We have shown that at small angles of falling of neutrons to a plane of converters it is possible to receive the efficiency of registration close to its theoretical limit. Therefore in the case of use of converters made from 157 isotope of gadolinium it is possible to receive the efficiency of registration 0.32, 0.35 and 0.65 for the cases of electron escapes in a forward hemisphere, a back one, and the total value of the efficiency, accordingly. At small angles (2-3$^0$) of falling of neutron flux the converters made from natural gadolinium can even compete to the converters made from 157 isotope of gadolinium.

During the design and manufacturing of detectors of the big areas the developers face a following problem, especially it relates to the case of use 157 isotope of gadolinium. The most optimal thickness of these converters lays between 0.5-3 microns, however this thickness of a foil do not possess the necessary mechanical durability and for their use a supporting film made from kapton or some other material with sufficient mechanical durability should be used. Efficiency of the complex converter designed of a supporting film made from kapton with the gadolinium converters layered on its surfaces from the both sides is considered also in this paper. It is shown, that at use of converters made from 157 isotope of gadolinium the basic absorption of neutrons is provided by the lobby converter, and the second converter is substantially shielded by the first one. The optimal thickness of converters is 1-2 microns. As concerns the converters made from natural gadolinium, the optimal way is the use of thicker converters (with the thickness more than 3 microns).

All calculations are performed for four fixed energies of neutrons, which correspond to the lengths of waves of 1, 1.8, 3 and 4 A$^0$, and for the energies of electrons of internal conversion and Auger electrons higher than 20 keV.

Authors are grateful to Dr. V.Dangendorf (PTB, Braunschweig, Germany) and Dr. B.Gebauer (HMI, Berlin, Germany) for stimulating discussions.

This research is conducted under the support of the International Science and Technology Center, Project T-1157.

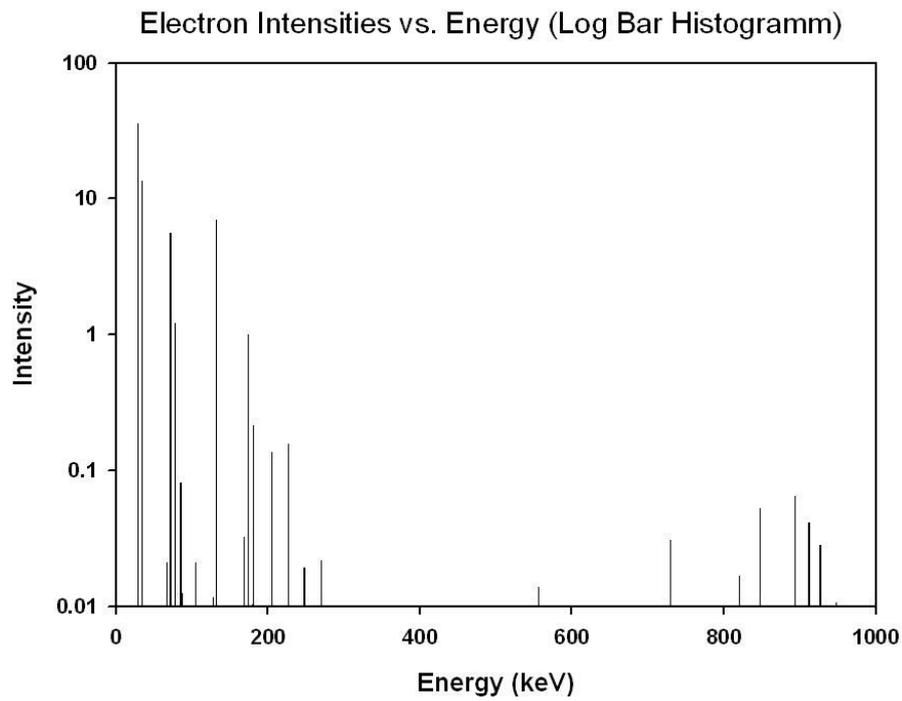

Fig. 1. Histogram of dependence of intensity of irradiated electrons on their energy in the reaction $^{157}Gd(n,\gamma)^{158}Gd$

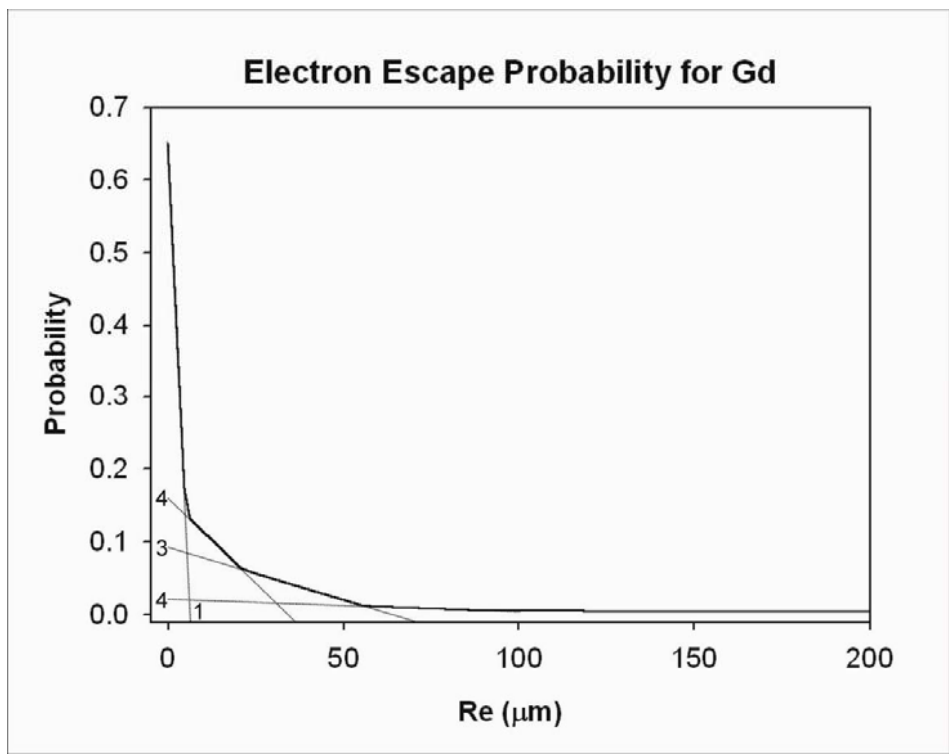

Fig. 2. Probability of escape of isotropic emitting secondary electrons from the gadolinium foils of a various thicknesses, taking into account their intensity and a solid angle

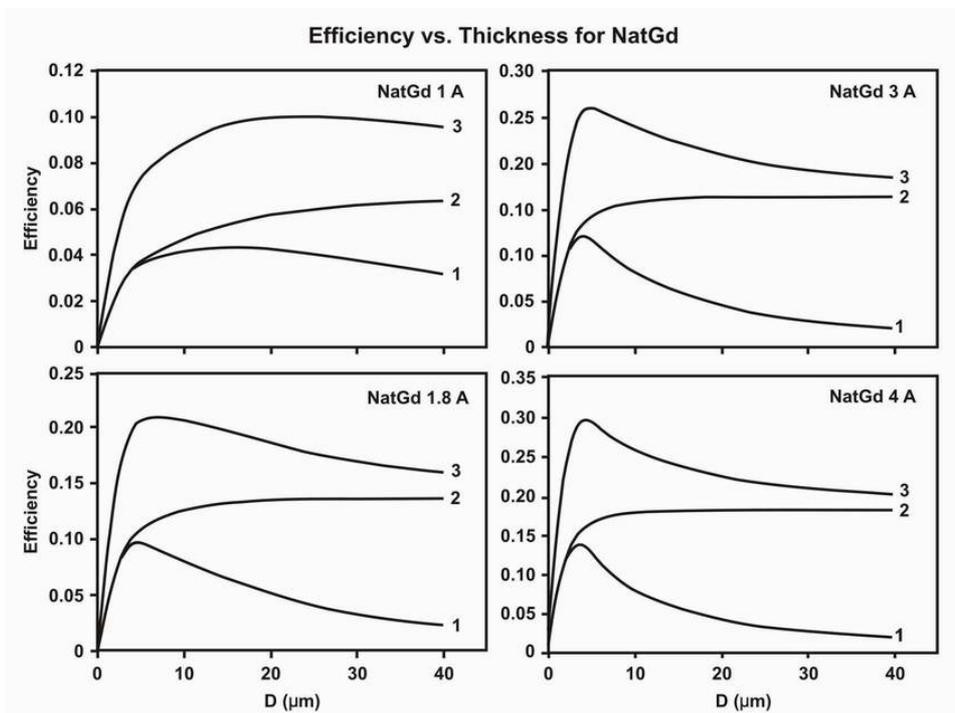

Fig. 3. Dependence of the efficiency of registration of thermal neutrons by the converters made from natural gadolinium depending on their thicknesses. Curves 1, 2 correspond to an electron escape in forward and back hemispheres accordingly, the curve 3 is their sum

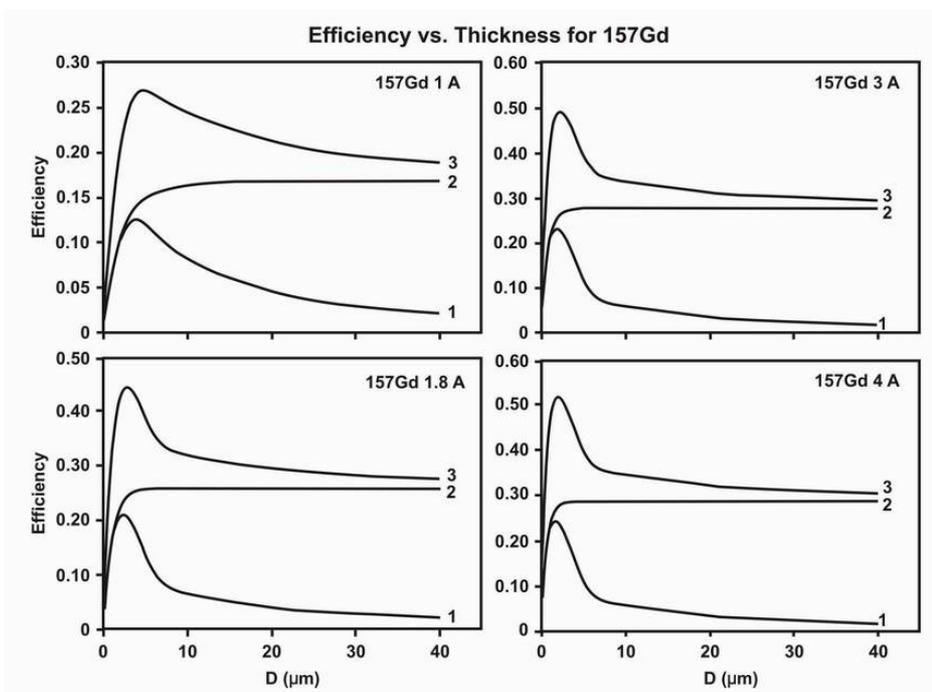

Fig. 4. Dependence of efficiency of registration of thermal neutrons by the converters made from [157]Gd depending on their thicknesses. Curves 1, 2 correspond to an electron escape in forward and back hemispheres accordingly, the curve 3 is their sum

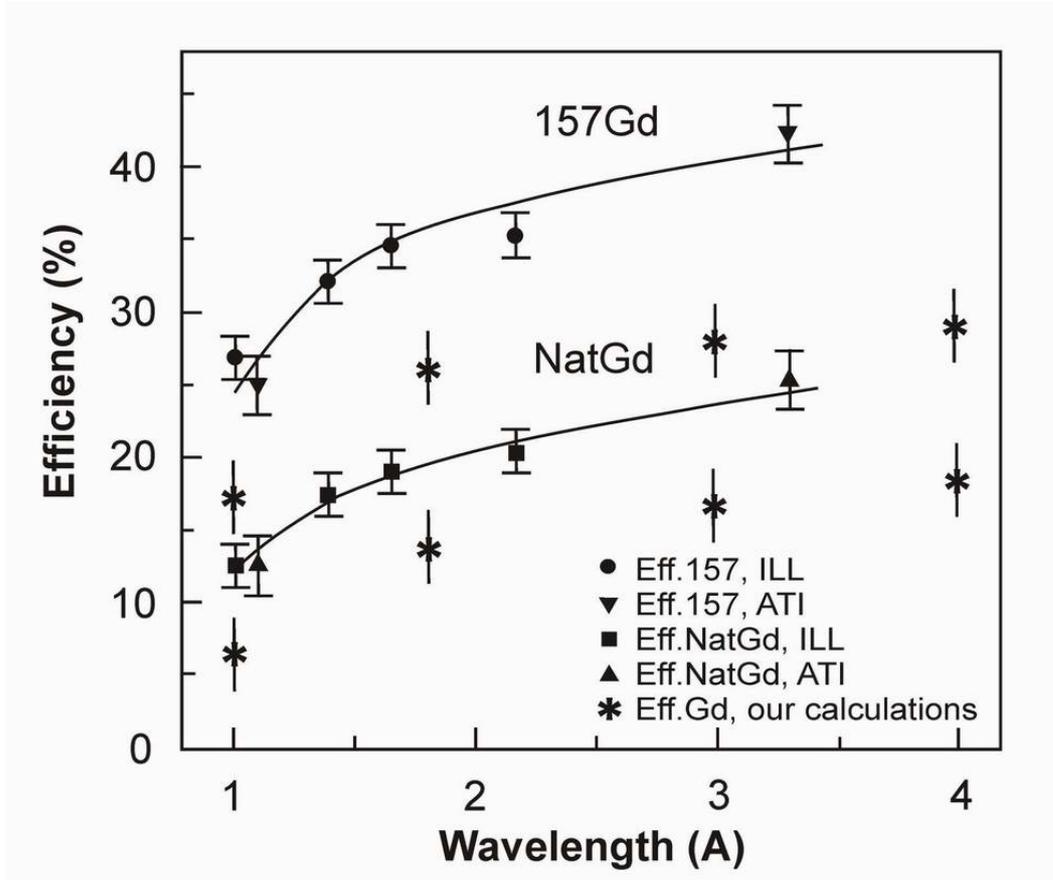

Fig. 5. Comparison of experimental data with the results of our calculations

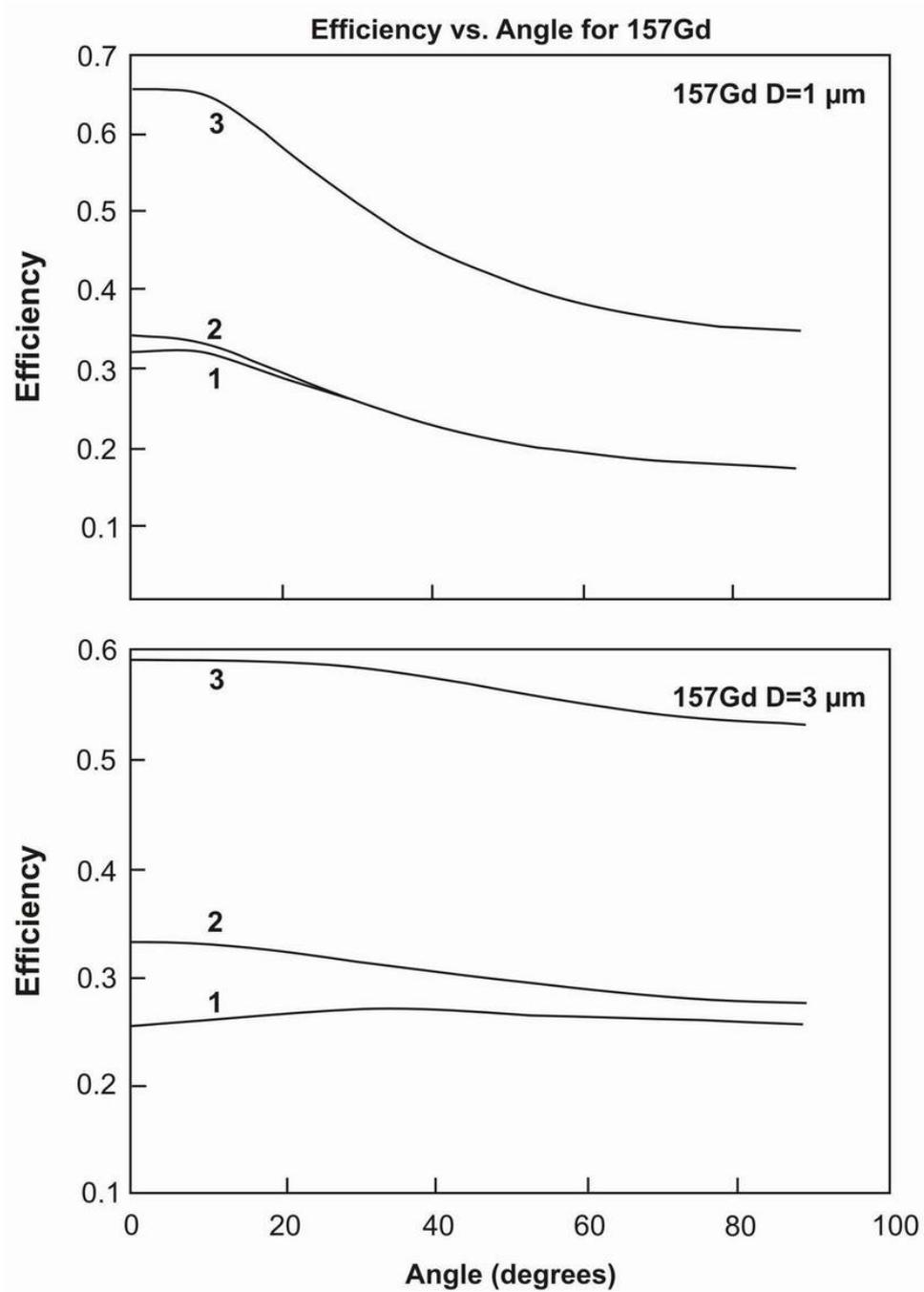

Fig. 6. Dependence of the efficiency of registration of neutrons with the wavelength 1.8 $A^0$ on a falling angle, for the thickness of converters ($^{157}$Gd) 1 and 3 microns. Curves 1 and 2 relate, accordingly, to an electron escape in forward and back hemispheres, the curve 3 is their sum

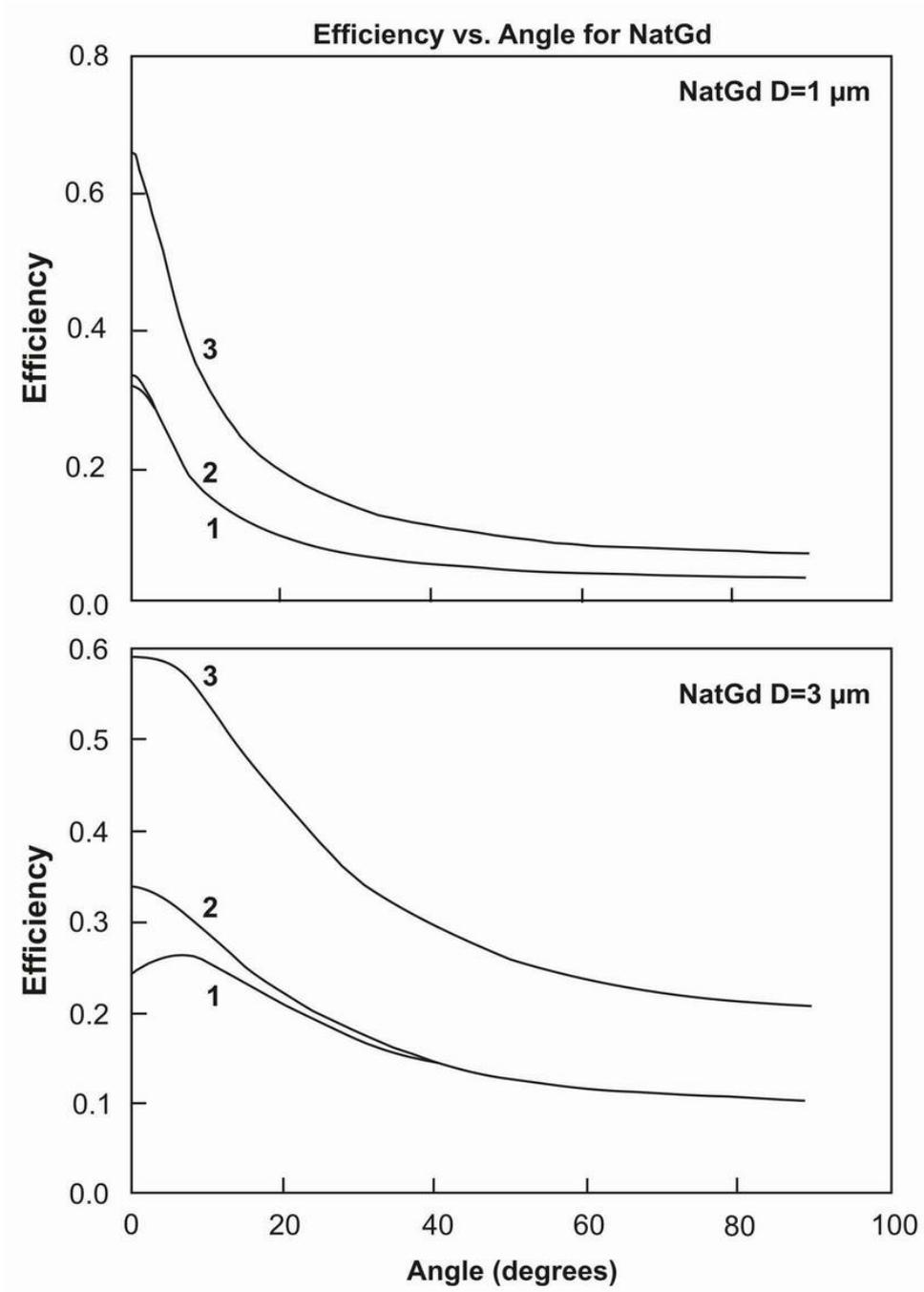

Fig. 7. Dependence of the efficiency of registration of neutrons with the wavelength of 1.8 $A^0$ on a falling angle of neutron flux, for the thickness of converters ($^{nat}Gd$) of 1 and 3 microns. Curves 1 and 2 relate, accordingly, to an electron escape in forward and back hemispheres, the curve 3 is their sum

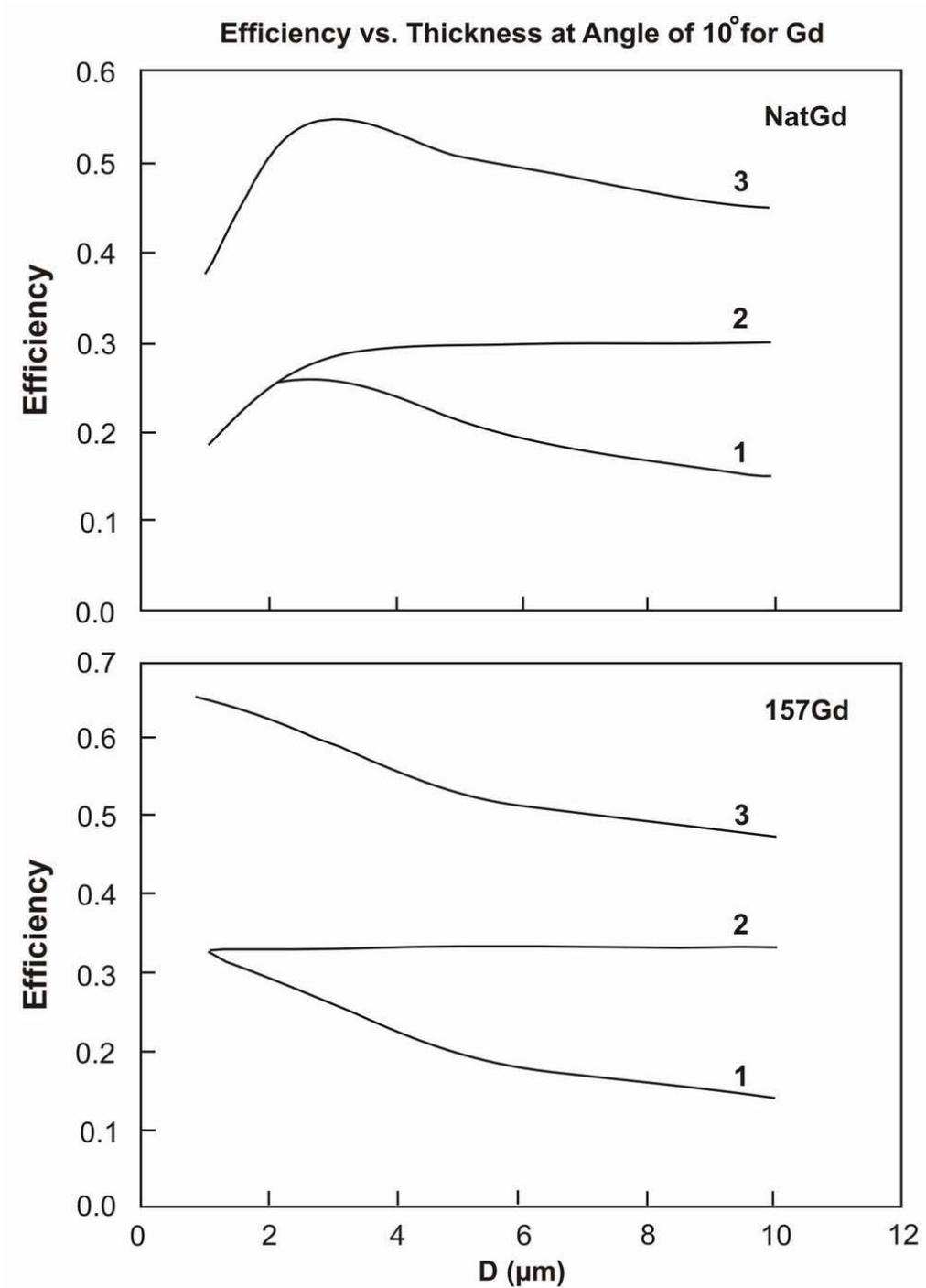

Fig. 8. Dependence of the efficiency of registration of thermal neutrons with wavelength of 1.8 A$^0$ on a thickness of the converter at falling of particles under the angle 10$^0$ to the plane of the converter

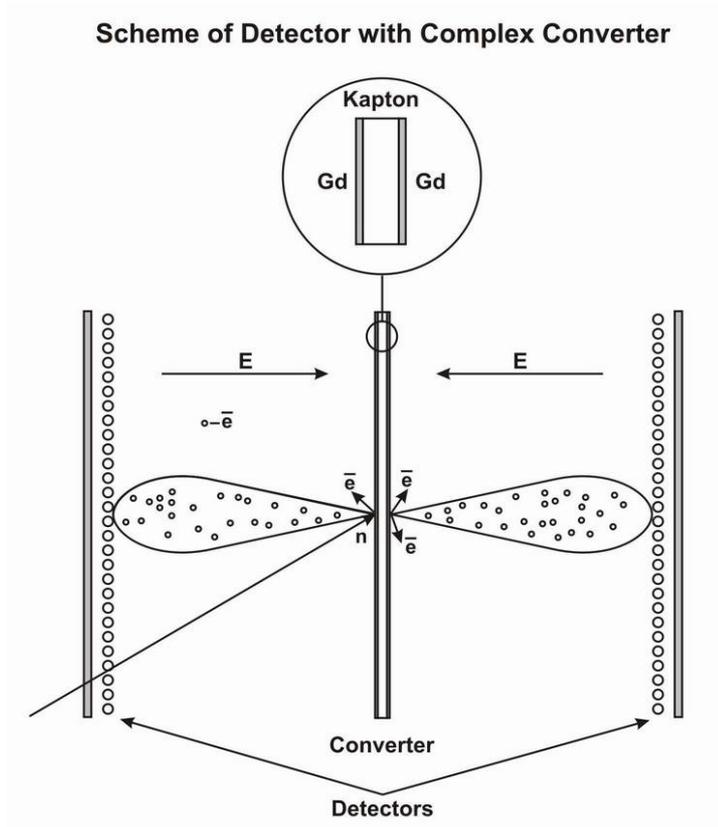

Fig. 9. The sketch of the detector with the complex converter consisting of the supporting film and two converters

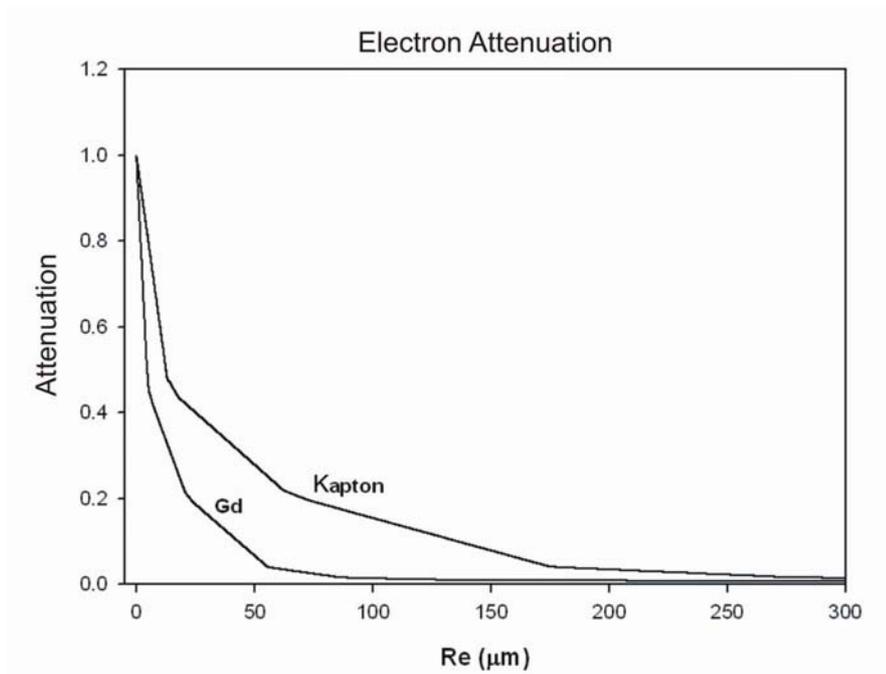

Fig. 10. Attenuation of an escape of electrons, formed in the reaction of radiating capture of thermal neutrons by gadolinium nuclei, from kapton and gadolinum

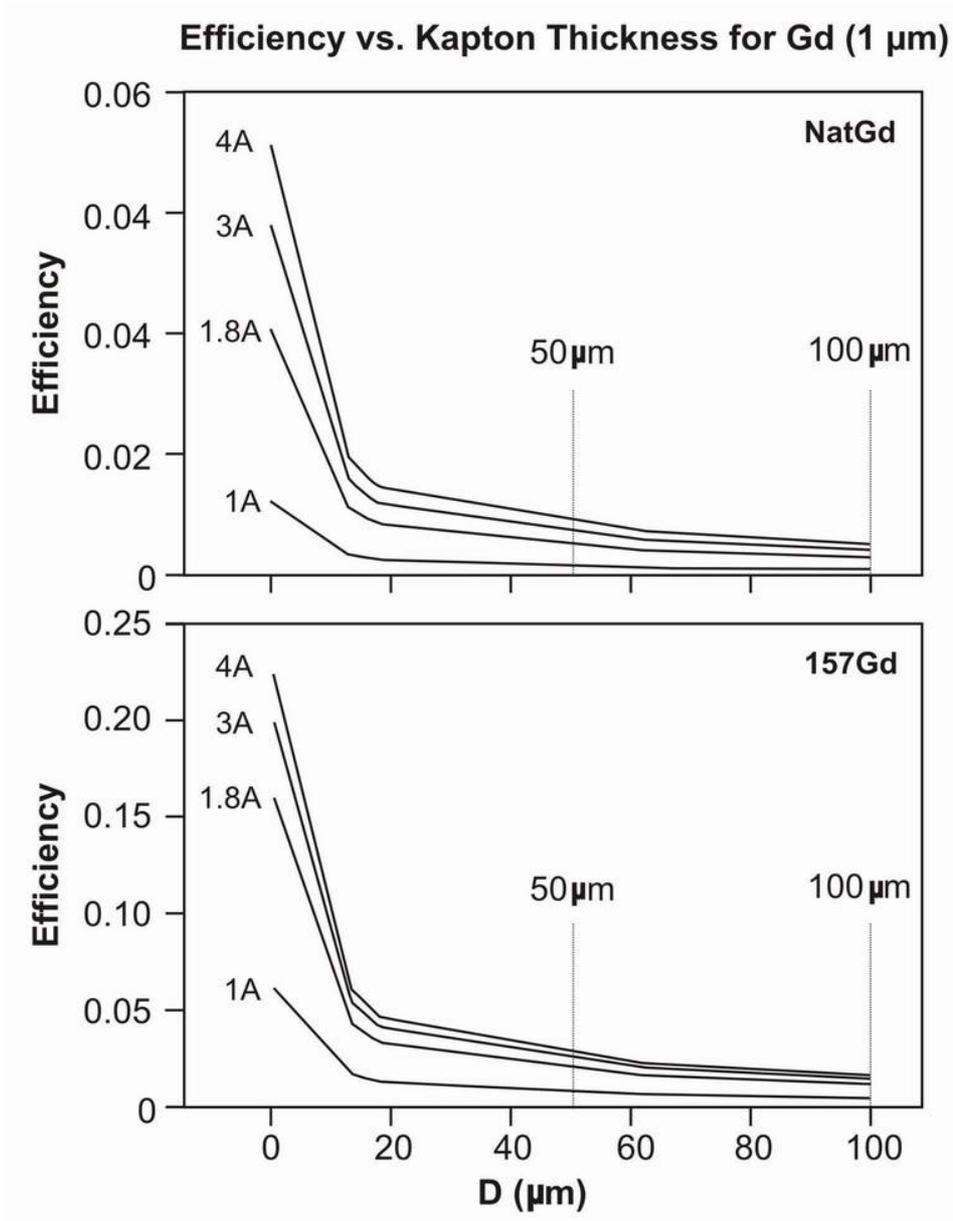

Fig. 11. Efficiency of complex converters made from thin gadolinium (1 micron) layered on a kapton substrate, depending on a thickness of a substrate. Different curves correspond to the different wavelengths of neutrons

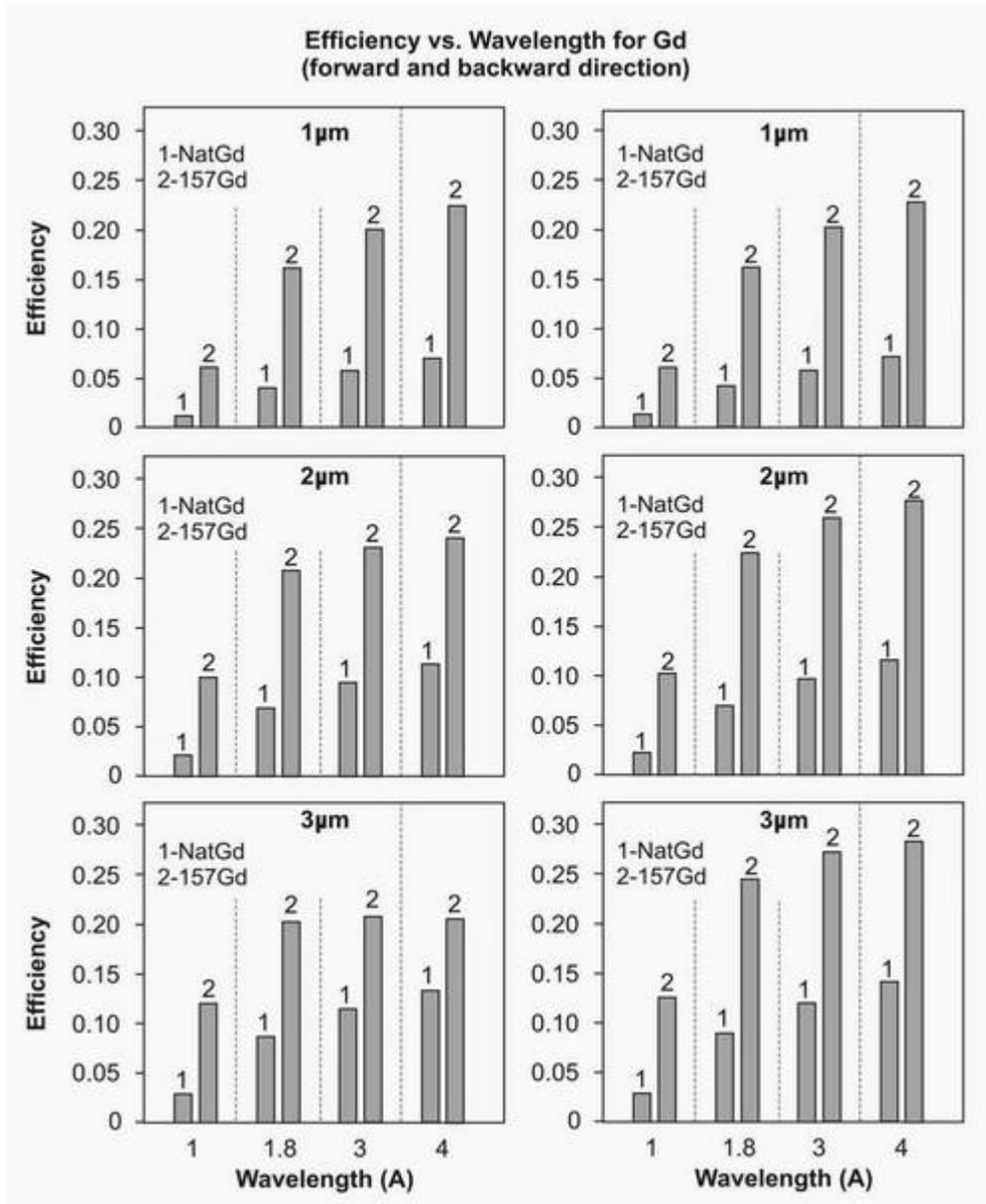

Fig. 12. Efficiency of registration of neutrons with the wavelengths of 1, 1.8, 3 and 4 $A^0$ for $^{nat}$Gd and $^{157}$Gd for converters with the thickness 1, 2 and 3 microns

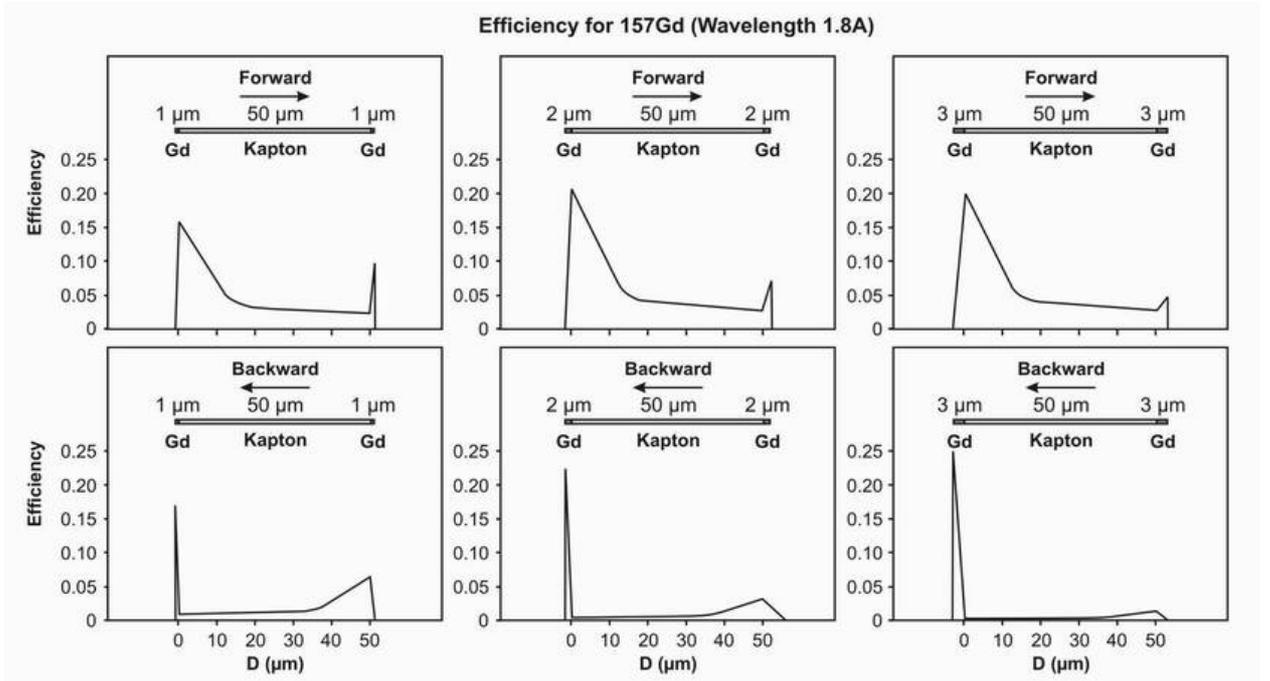

Fig. 13. Efficiency of the complex converter for the both cases of electron escape in the forward and back hemispheres. Thickness of the converter made from $^{157}$Gd is varied as 1, 2 and 3 microns, thickness of kapton film is 50 microns. Neutron wavelength is 1.8A$^0$.

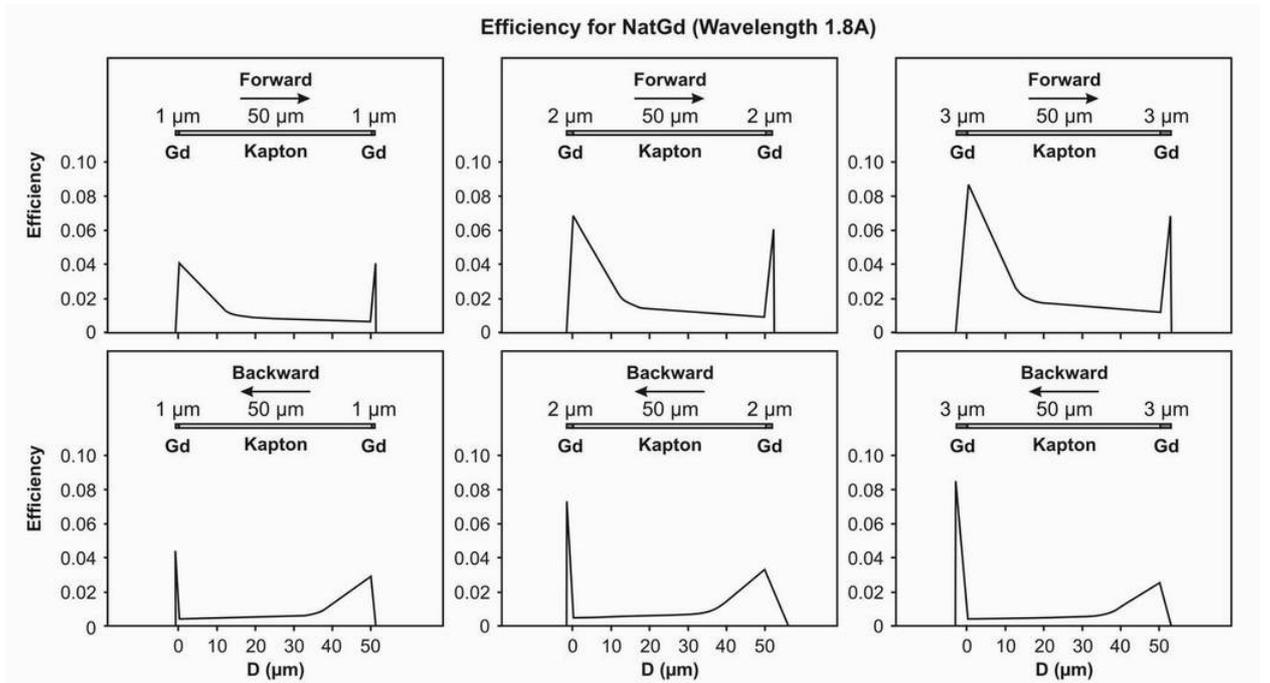

Fig. 14. Efficiency of the complex converter for the case of electron escape in forward and back hemispheres. Thickness of the converter made from $^{nat}$Gd is varied as 1, 2 and 3 microns, thickness of kapton film is 50 microns. Neutron wavelength is 1.8 A$^0$.